\documentstyle[12pt]{article}
\newlength{\extraspace}
\newlength{\extraspaces}
\newcommand{\be}{\begin{equation}
\addtolength{\abovedisplayskip}{\extraspaces}
\addtolength{\belowdisplayskip}{\extraspaces}
\addtolength{\abovedisplayshortskip}{\extraspace}
\addtolength{\belowdisplayshortskip}{\extraspace}}
\newcommand{\ee}{\end{equation}}

\newcommand{\ba}{\begin{eqnarray}
\addtolength{\abovedisplayskip}{\extraspaces}
\addtolength{\belowdisplayskip}{\extraspaces}
\addtolength{\abovedisplayshortskip}{\extraspace}
\addtolength{\belowdisplayshortskip}{\extraspace}}
\newcommand{\ea}{\end{eqnarray}}

\newcommand{\nonu}{\nonumber \\[.5mm]}
\newcommand{\A}{&\!\!\!}

\newcommand{\newsection}[1]{
\vspace{7mm} \pagebreak[3] \addtocounter{section}{1}
\setcounter{subsection}{0} \setcounter{footnote}{0}
\begin{center}
{\large {\bf \thesection. #1}}
\end{center}
\nopagebreak
\medskip
\nopagebreak \hspace{3mm}}

\begin{document}

\begin{center}
{{\bf Self-dual Lorentzian Wormholes and Energy in Teleparallel
Theory of Gravity }\footnote{Mathematics Department, Faculty of
Science, Ain
Shams University, Cairo, Egypt
}}
\end{center}
\centerline{ Gamal G.L. Nashed}

\bigskip

\centerline{\it Centre for Theoretical Physics, The British
University in Egypt,
 El-Sherouk City,}
\centerline{{\it Misr - Ismalia Desert Road, Postal No. 11837,
P.O. Box 43, Egypt.}} \centerline{ Gamal G.L. Nashed}

\bigskip
 \centerline{ e-mail:nashed@bue.edu.eg}

\hspace{2cm}
\\
\\
\\
\\
\\
\\
\\
\\

Two spherically symmetric, static Lorentzian wormholes are
obtained in tetrad theory of gravitation as a solution of the
equation $\rho=\rho_t=0$, where $\rho=T_{ij}u^iu^j,\
\rho_t=(T_{ij}-\frac{1}{2}Tg_{ij})u^iu^j$ and $u^iu_i=-1$. This
equation characterizes a class of spacetime which are ``self-dual"
(in the sense of electrogravity duality). The obtained solutions
are characterized by two-parameters $k_1,\ \ k_2$  and have a
common property that they reproduce the same metric spacetime.
This metric  is the static Lorentzian wormhole and it includes the
Schwarzschild black hole. Calculating the energy content of these
tetrad fields using the {\it superpotential method given by M\o
ller in the context of teleparallel spacetime} we find that $E=m$
or $2m$ which does not depend on the two parameters $k_1$ and
$k_2$ characterize the wormhole.
\newpage
\begin{center}
\newsection{\bf Introduction}
\end{center}

It was recognized by Flamm \cite{Fl} in (1916) that our universe
may not be simply connected, there may exist handles or tunnels
now called wormholes, in the spacetime topology linking widely
separated regions of our universe or even connected us with
different universes altogether. traversable Lorentzian wormholes
"which have no horizons, allowing two-way passage through them",
and were especially stimulated by the pioneering works of Morris,
Thorne and Yurtsever \cite{MT,MTY}, where static, spherically
symmetric Lorentzian wormholes were defined and considered to be
an exciting possibility for constructing time machine models with
these exotic objects, for backward time travel \cite{Njd,Vi,Kp}.

Morris and Thorne (MT) wormholes are static and spherically
symmetric and connect asymptotically flat spacetimes.  The metric
of this wormhole is given by \be
ds^2=-e^{2\Phi(r)}dt^2+\displaystyle{dr^2 \over 1-b(r)/r}+r^2(
d\theta^2+ \sin^2 \theta d\phi^2)\; , \ee where $\Phi(r)$ being
the redshift function and $b(r)$ is the shape function. The shape
function describes the spatial shape of the wormhole when viewed.
The metric (1) is spherically symmetric and static.   The
coordinate $r$ is nonmonotonic in that it decreases from $+
\infty$ to a minimum value $b_0$, representing the location of the
throat of the wormhole, and then it increases from $b_0$ to $+
\infty$. This behavior of the radial coordinate reflects the fact
that the wormhole connects two separate external universes. At the
throat  $r=b=b_0$, there is a coordinate singularity where the
metric coefficient $g_{rr}$ becomes divergent but the radial
proper distance \be l(r)= \pm{\int_{b_0}}^r \displaystyle{ dr
\over \sqrt{1-b(r)/r}},\ee must be required to be finite
everywhere \cite{Roa}. At the throat, $l(r)=0$, while $l(r)<0$ on
the left side of the throat and $l(r)>0$ on the right side. For a
wormhole to be traversable it must have no horizon which implies
that $g_{tt}$ must never allowed to be vanish, i.e., $\Phi(r)$
must be finite everywhere.

Traversable Lorentzian have been in vogue ever since Morris, Thorn
and Yurtsever  \cite{MTY} came up with the exciting possibility of
constructing time machine models with these exotic objects. (MT)
paper demonstrated that the matter required to support such
spacetimes necessarily violates the null energy condition.
Semiclassical calculations based on techniques of quantum fields
in curved spacetime, as well as an old theorem of Epstein et. al.
\cite{EGY}, raised hopes about generation of such spacetimes
through quantum stresses.

There have been innumerable attempts at solving the "exotic matter
problem" in wormhole physics in the last few years \cite{Vi,VH}.
Alternative theories of gravity \cite{Hd} evolving wormhole
spacetimes \cite{Ks}$\sim$ \cite{HV1} with varying definitions of
the throat have been tried out as possible avenues of resolution.

 M\o ller modified general relativity by constructing a new
field theory in Weitzenb$\ddot{o}$ck spacetime \cite{Mo8}. The aim
of this theory was to overcome the problem of energy-momentum
complex that appears in the Riemannian spacetime \cite{M2}.  The
field equations in this new theory were derived from a Lagrangian
which is not invariant under local tetrad rotation. S$\acute{a}$ez
\cite{Se} generalized M\o ller theory into a scalar tetrad theory
of gravitation.  Meyer \cite{Me} showed that M\o ller theory is a
special case of Poincar$\acute{e}$ gauge theory \cite{HS8,HNV}.

The tetrad theory  of gravitation  based on the geometry of
absolute parallelism \cite{PP}$\sim$\cite{AGP} can be considered
as the closest alternative to general relativity, and it has a
number of attractive features both from the geometrical and
physical viewpoints. Absolute parallelism is naturally formulated
by gauging spacetime translations and underlain by the
Weitzenb$\ddot{o}$ck spacetime, which is characterized by the
metric condition and by the vanishing of the curvature tensor
using the affine connection resulting from the geometry of the
Weitzenb$\ddot{o}$ck spacetime. Translations are closely related
to the group of general coordinate transformations which underlies
general relativity. Therefore, the energy-momentum tensor
represents the matter source in the field equation for the
gravitational field just like in general relativity.

It is the aim of the present work to derive a spherically
symmetric solutions in the tetrad theory of gravitation. To do so
we first begin with a tetrad having spherical symmetry with three
unknown functions of the radial coordinate \cite{Ro}. Applying
this tetrad to the field equations of M\o ller's theory we obtain
a set of non linear partial differential equations.  We propose a
specific restriction on the form of the stress-energy that when
solving the non linear partial differential equations,
automatically leads to a class of wormhole solutions. The
characterization of our class of self-dual wormholes is \be
\rho=\rho_t=0,\ee where $\rho$ and $\rho_t$ are respectively the
energy density measured by a static observer and the convergence
density felt by a timelike congruence. Indeed Eq. (3)
characterizes a class of ``self-dual" wormhole spacetimes which
contains the Schwarzschild solution. The notation of duality
involves the interchange of the active and passive electric parts
of the Riemann tensor (termed as electrogravity duality)
\cite{DKV}. Electrogravity duality essentially implies the
interchange of the Ricci and Einstein tensors. For vanishing Ricci
scalar tensor the Ricci and Einstein tensors become equal and the
corresponding solution could be called ``self-dual" in this sense.

The weak energy condition says that the energy density of any
system at any point of spacetime for any timelike observer is
positive (in the frame of the matter this amounts to $\rho>0$ and
$\rho+p \geq0 $). When the observer moves at the speed of light it
has a well defined limit, called the null energy condition (in the
frame of matter $\rho+p \geq0 $) \cite{JLO}.

Under the duality transformation, $\rho$ and $\rho_t$ are
interchanged indicating invariance of Eq. (3). Since energy
densities vanish and yet the spacetime is not entirely empty, the
matter distribution would naturally have to be exotic (violating
the energy condition) \cite{DKV}. Physically, the existence of
such spacetimes might be doubted because of this violation of the
weak and null energy conditions. Analogous to the
spatial-Schwarzschild wormhole, for which $g_{00}=-1$ and
$g_{11}=(1-\frac{2m}{r})^{-1}$, these spacetimes have zero energy
density but nonzero pressures \cite{DKV}. The
spatial-Schwarzschild wormhole is one specific particular solution
of the equations $\rho=0, \ \rho_t=0$. The central problem of the
traversable wormhole is connected with the unavoidable violation
of the null energy condition. This means that the matter which
should be a source of this object has to possess some exotic
properties. For this reason the traversable wormhole cannot be
represented as a self-consistence solution of Einstein's equations
with the usual classical matter as a source because the usual
matter is sure to satisfy all the energy conditions \cite{Kn}.
Therefore, it is a basic fact for the construction of the
traversable wormholes that the null energy condition has to be
violated \cite{JLO}.

It is our aim to discuss the physical properties of the
spherically symmetric wormholes obtained. In section 2 a brief
survey of M\o ller's tetrad theory of gravitation is presented.
The exact solutions of the set of  non linear partial differential
equations are given in section 3. In section 4 the energy content
of these solutions are given. Discussion and conclusion of the
obtained results are given in section 5.
\newpage
\newsection{M\o ller's  tetrad theory of gravitation}

In a spacetime with absolute parallelism the parallel vector
fields ${e_i}^\mu$ define the nonsymmetric affine connection \be
{\Gamma^\lambda}_{\mu \nu} \stackrel{\rm def.}{=} {e_i}^\lambda
{e^i}_{\ \mu, \ \nu}, \ee where ${e^i}_{\mu, \ \nu}=\partial_\nu
{e^i}_{\mu}$. The curvature tensor defined by
${\Gamma^\lambda}_{\mu \nu}$ is identically vanishing, however.

M\o ller's constructed a gravitational theory based on
 this spacetime. In this
theory the field variables are the 16 tetrad components
${e_i}^\mu$, from which the metric tensor is derived by \be g^{\mu
\nu} \stackrel{\rm def.}{=} \eta^{i j} {e_i}^\mu {e_j}^{\nu}, \ee
where $\eta^{i j}$ is the Minkowski metric $\eta_{i j}=\textrm
{diag}(+1\; ,-1\; ,-1\; ,-1).$

 We note that, associated with any tetrad field ${e_i}^\mu$ there
 is a metric field defined
 uniquely by (5), while a given metric $g^{\mu \nu}$ does not
 determine the tetrad field completely; for any local Lorentz
 transformation of the tetrads ${b_i}^\mu$ leads to a new set of
 tetrads which also satisfy (5).
  The Lagrangian ${\it L}$ is an invariant constructed from
$\gamma_{\mu \nu \rho}$ and $g^{\mu \nu}$, where $\gamma_{\mu \nu
\rho}$ is the contorsion tensor given by \be \gamma_{\mu \nu \rho}
\stackrel{\rm def.}{=} \eta^{ij}e_{i \ \mu }e_{j \nu; \ \rho}, \ee
where the semicolon denotes covariant differentiation with respect
to Christoffel symbols. The most general Lagrangian density
invariant under the parity operation is given by the form
\cite{Mo8} \be {\cal L} \stackrel{\rm def.}{=} \sqrt{-g} \left(
\alpha_1 \Phi^\mu \Phi_\mu+ \alpha_2 \gamma^{\mu \nu \rho}
\gamma_{\mu \nu \rho}+ \alpha_3 \gamma^{\mu \nu \rho} \gamma_{\rho
\nu \mu} \right), \ee where \be g \stackrel{\rm def.}{=} {\rm
det}(g_{\mu \nu}),
 \ee
 and
$\Phi_\mu$ is the basic vector field defined by \be \Phi_\mu
\stackrel{\rm def.}{=} {\gamma^\rho}_{\mu \rho}. \ee Here
$\alpha_1, \alpha_2,$ and $\alpha_3$ are constants determined by
M\o ller such that the theory coincides with general relativity in
the weak fields:

\be \alpha_1=-{1 \over \kappa}, \qquad \alpha_2={\lambda \over
\kappa}, \qquad \alpha_3={1 \over \kappa}(1-2\lambda), \ee where
$\kappa$ is the Einstein constant and  $\lambda$ is a free
dimensionless parameter\footnote{Throughout this paper we use the
relativistic units, $c=G=1$ and
 $\kappa=8\pi$.}. The same
choice of the parameters was also obtained by Hayashi and Nakano
\cite{HN}.

M\o ller applied the action principle to the Lagrangian density
(7) and obtained the field equation in the form \be G_{\mu \nu}
+H_{\mu \nu} = -{\kappa} T_{\mu \nu}, \qquad K_{\mu \nu}=0, \ee
where the Einstein tensor $G_{\mu \nu}$ is the Einstein tensor
defined by \[G_{\mu \nu} \stackrel{\rm def.}{=} R_{\mu
\nu}(\{\})-\frac{1}{2}g_{\mu \nu} R(\{\}) ,\] where $R_{\mu
\nu}(\{\})$ and $R(\{\})$ are the Ricci tensor and Ricci scalar.
$H_{\mu \nu}$ and $K_{\mu \nu}$ are given by \be H_{\mu \nu}
\stackrel{\rm def.}{=} \lambda \left[ \gamma_{\rho \sigma \mu}
{\gamma^{\rho \sigma}}_\nu+\gamma_{\rho \sigma \mu}
{\gamma_\nu}^{\rho \sigma}+\gamma_{\rho \sigma \nu}
{\gamma_\mu}^{\rho \sigma}+g_{\mu \nu} \left( \gamma_{\rho \sigma
\tau} \gamma^{\tau \sigma \rho}-{1 \over 2} \gamma_{\rho \sigma
\tau} \gamma^{\rho \sigma \tau} \right) \right],
 \ee
and \be K_{\mu \nu} \stackrel{\rm def.}{=} \lambda \left[
\Phi_{\mu,\nu}-\Phi_{\nu,\mu} -\Phi_\rho \left({\gamma^\rho}_{\mu
\nu}-{\gamma^\rho}_{\nu \mu} \right)+ {{\gamma_{\mu
\nu}}^{\rho}}_{;\rho} \right], \ee and they are symmetric and skew
symmetric tensors, respectively.

M\o ller assumed that the energy-momentum tensor of matter fields
is symmetric. In the Hayashi-Nakano theory, however, the
energy-momentum tensor of spin-$1/2$ fundamental particles has
non-vanishing antisymmetric part arising from the effects due to
intrinsic spin, and the right-hand side of antisymmetric field
equation  (11) does not vanish when we take into account the
possible effects of intrinsic spin.

It can be shown \cite{HS} that the tensors, $H_{\mu \nu}$ and
 $K_{\mu \nu}$, consist of only those terms which are linear or quadratic
in the axial-vector part of the torsion tensor, $a_\mu$, defined
by \be a_\mu \stackrel{\rm def.}{=} {1 \over 3} \epsilon_{\mu \nu
\rho \sigma} \gamma^{\nu \rho \sigma}, \qquad where \qquad
\epsilon_{\mu \nu \rho \sigma} \stackrel{\rm def.}{=} \sqrt{-g}
\delta_{\mu \nu \rho \sigma}, \ee where $\delta_{\mu \nu \rho
\sigma}$ being completely antisymmetric and normalized as
$\delta_{0123}=-1$. Therefore, both $H_{\mu \nu}$ and $F_{\mu
\nu}$ vanish if the $a_\mu$ is vanishing. In other words, when the
$a_\mu$ is found to vanish from the antisymmetric part of the
field equations, (11), the symmetric part of Eq. (11) coincides
with the Einstein field equation in teleparallel equivalent of
general relativity.
\newsection{Spherically Symmetric Solutions}

Let us begin with the tetrad \cite{Ro} (lines are labelled by $l$
and columns by $\mu$) \be \left({e_l}^\mu \right)= \left( \matrix{
A & Dr & 0 & 0 \vspace{3mm} \cr 0 & B \sin\theta \cos\phi &
\displaystyle{B \over r}\cos\theta \cos\phi
 & -\displaystyle{B \sin\phi \over r \sin\theta} \vspace{3mm} \cr
0 & B \sin\theta \sin\phi & \displaystyle{B \over r}\cos\theta
\sin\phi
 & \displaystyle{B \cos\phi \over r \sin\theta} \vspace{3mm} \cr
0 & B \cos\theta & -\displaystyle{B \over r}\sin\theta  & 0 \cr }
\right), \ee where {\it A}, {\it D}, {\it B}, are functions of the
redial coordinate $r$. The associated metric of the tetrad (14)
has the form \be ds^2=-\displaystyle{B^2-D^2r^2 \over
A^2B^2}dt^2-2\displaystyle{Dr \over AB^2}drdt+\displaystyle{1
\over B^2}dr^2+\displaystyle{r^2 \over B^2}(d\theta^2+ \sin^2
\theta d\phi^2). \ee As is clear from (16) that there is a cross
term which can be eliminated by performing the coordinate
transformation \cite{Wm} \be dT=dt+{ADr \over B^2-D^2r^2} dr, \ee
using the transformation (17) in the tetrad (15) we obtain \be
\left({e_l}^\mu \right)= \left( \matrix{ \displaystyle {{\cal A}
\over 1-{\cal D}^2R^2} & (R{\cal D}-R^2{\cal D}{\cal B}') & 0 & 0
\vspace{3mm} \cr \displaystyle{{\cal A}{\cal D}R \sin\theta
\cos\phi \over 1-{\cal D}^2R^2}& (1-R{\cal B}')\sin\theta \cos\phi
& \displaystyle{\cos\theta \cos\phi \over R} &
-\displaystyle{\sin\phi \over R \sin\theta} \vspace{3mm} \cr
 \displaystyle{{\cal A}{\cal D}R \sin\theta \sin\phi \over
1-{\cal D}^2R^2} & (1-R{\cal B}') \sin\theta \sin\phi &
\displaystyle{\cos\theta \sin\phi \over R} &
\displaystyle{\cos\phi \over R \sin\theta} \vspace{3mm} \cr
\displaystyle{{\cal A}{\cal D}R \cos\theta  \over 1-{\cal D}^2R^2}
& (1-R{\cal B}') \cos\theta & -\displaystyle{\sin\theta  \over R}
& 0 \cr } \right), \ee where ${\cal A}$, ${\cal D}$ and ${\cal B}$
are now unknown functions of the new radial coordinates $R$ which
is connected to the old redial coordinate $r$ through the
transformation defined by \be R=\displaystyle{r \over B}, \qquad
with \qquad {\cal B}'=\displaystyle{d{\cal B}(R) \over dR}.\ee

We are interested in finding some special solutions to the partial
non linear differential equations resulting from applying the
tetrad (18) to the field equations (11).

\underline {The First Solution}

If the unknown function ${ \cal D(R)}=0$, then the resulting
partial non linear differential equations  have the form
 \ba
  \rho(R) \A= \A -\displaystyle{1 \over R \kappa}\left\{ 3\,  {{\cal B}'}^2 R-2
\,R{\cal B}'' \left( R \right) +2\,{R}^{2} {\cal B}' \left( R
\right) {\cal B}'' \left( R \right) -4\,{\cal B}' \left( R
\right)\right\} , \nonu
 \tau(R) \A= \A {1 \over \kappa R {\cal A}}\left\{2\,  {{\cal B}'} {\cal
 A}-4R\,  {{\cal A}'} {\cal
 B}'+2\,  {{\cal A}'}+2R^2\,  {{\cal A}'} {{\cal
 B}'}^2-R{{\cal A}}{{\cal B}'}^2\right\} ,\nonu
 p(R) \A=\A  \kappa {T^3}_3= {1-R {{\cal B}'}^2 \over \kappa R
{\cal A}^2}\left\{2R^2\,  {{\cal A}'}^2 {\cal
 B}'- {\cal A} [R^2{\cal A}' {\cal B}']'+{\cal A}^2 [R{\cal B}']'+{\cal A} [R{\cal A}']'-
 2R {\cal A}''\right\},
 \ea
where \[\rho(R)={T^0}_0, \qquad \qquad \tau(R)={T^1}_1, \qquad
\qquad p(R)={T^2}_2={T^3}_3,\] with $\rho(R)$ being the {\it
energy density}, $\tau(R)$ is the {\it radial pressure} and $p(R)$
is {\it the tangential pressure}. (Note that $\tau(R)$ as defined
above is simply the radial pressure $p_r$, and differs by a minus
sign from the conventions in \cite{MT,Vi}.)

Eqs. (20) can be solved to take the form \be {\cal
A(R)}=\displaystyle{1 \over k_1+k_2\sqrt{1-\displaystyle{2m \over
R}}}, \quad \quad {\cal
B(R)}=\ln\left\{R\left(R-m+R\sqrt{1-\displaystyle{2m \over
R}}\right)\right\}-2\sqrt{1-\displaystyle{2m \over R}}, \ee where
$k_1$ and $k_2$ are constants of integration. The associated
Riemannian metric of solution (21) takes the form \be ds^2=
-\eta_1(R) dT^2 +{dR^2 \over \eta_2(R)} +R^2d\Omega^2, \quad where
\quad \eta_1(R)=\left(k_1+k_2\sqrt{1-\displaystyle{2m \over R}}
\right)^2, \quad  \eta_2(R)=\left({1-\displaystyle{2m \over R}}
\right), \ee with ${d\Omega^2=d\theta^2+\sin^2\theta d\phi^2}$.
\newpage
 \underline {The Second Solution}

If the unknown function ${ \cal B(R)}=1$, then the resulting
partial non linear differential equations  have the form
 \ba
  \rho(R) \A= \A  \displaystyle{{\cal D} \over  \kappa}\left\{ 3\,  {\cal D}+2
\,R{\cal D}' \right\}, \nonu
 \tau(R) \A= \A -\displaystyle{{\cal D}(1-R {{\cal D}'}^2) \over {\cal A}^2 \kappa}\left\{ 3\,  {\cal D}+2
\,R{\cal D}' \right\} ,\nonu
 p(R) \A=\A  \kappa {T^3}_3= {1 \over \kappa R
{\cal A}^2}\Biggl\{R^3 {\cal A}^2 {\cal D}\, {{\cal D}''}+6R^2
{\cal A}^2 {\cal D} {\cal D}'+3R{\cal A}^2 {\cal D}^2+{\cal
A}{\cal A}'- 4R^2 {\cal A}{\cal D}^2{\cal A}'\nonu
\A \A -(2R{{\cal A}'}^2-R{\cal A}{\cal A}'')(1-R^2{\cal
D}^2)-3R^3{\cal A} {\cal D} {\cal A}'{\cal D}'+R^3{\cal A}^2{{\cal
D}'}^2\Biggr\}.
 \ea
 Eqs. (23) can be solved to take the form  \be {\cal
A(R)}=\displaystyle{1 \over k_2+\displaystyle{k_1 \over
\sqrt{1-\displaystyle{2m \over R}}}}, \quad \quad  {\cal
D(R)}=\sqrt{\displaystyle{2m \over R^3}}. \ee  Using (21) or (24)
in (11) we can get the components of the energy-momentum tensor
turn out to have the form
 \be \rho(R)= 0,\qquad
 \tau(R)  = -\displaystyle{1 \over
 \kappa}\displaystyle{\left[2mk_1 \over
 R^3\left(k_1+k_2\sqrt{1-\displaystyle{2m \over R}}\right)
 \right]},\qquad
  p(R) = \displaystyle{1 \over
\kappa}\left[\displaystyle{mk_1 \over
R^3\left(k_1+k_2\sqrt{1-\displaystyle{2m \over R}}\right)}
\right]. \ee The weak energy \be \rho\geq 0, \qquad  \rho+\tau
\geq 0, \qquad  \rho+p \geq 0, \ee  and  null energy conditions
\be \rho+\tau \geq 0, \qquad \rho+p \geq 0\ee  are both violated
as is clear from (25). The violation of the energy condition stems
from the violation of the inequality $\rho+\tau \geq 0$.

The associated Riemannian space of solution (24) has the form
(22). If one replacing $k_2$ by $-k_2$ at the above solutions,
((21) or (24)), the resulting form will also be a solution to the
non linear partial differential equations ((20) and (23)). The
Ricci scalar tensor vanishing, i.e.,
\[R(\{\})=0, \] for solutions (21) and
(24).  The metric (22) makes sense only for $R\geq 2m$ so to
really make the wormhole explicit one needs two conditions
patches\[ R_1 \in (2m ,\infty), \qquad \qquad R_2 \in (2m
,\infty),\] which we then have to sew together at $R=2m$.

Transforming the metric (22) to the isotropic coordinate using the
transformation \be R \rightarrow {\bar R}\left(1+\displaystyle{m
\over 2{\bar R}}\right)^2.\ee Using (28) in (22), then the
transformed metric will have the form \be ds^2= -\eta_1({\bar R})
dT^2 +\eta_2({\bar R})\left[d{\bar R}^2+{\bar R}^2 d
\Omega^2\right], \ee where $\eta_1({\bar R})$ and $\eta_2({\bar
R})$ have the form \be \eta_1({\bar
R})=\left(k_1+k_2\left[\displaystyle{{1-\displaystyle{m \over
2{\bar R}}} \over {1+\displaystyle{m \over 2{\bar
R}}}}\right]\right)^2 , \qquad \qquad \eta_2({\bar
R})=\left({1+\displaystyle{m \over 2{\bar R}}} \right)^4,\ee and
the energy momentum tensor given by Eq. (25) takes the form \ba
\tau(R) \A =\A -\displaystyle{128 k_1m \over
 \kappa}\displaystyle{\left[1 \over
 {\bar R}^2\left(1+ \displaystyle{m \over 2{\bar
R}}\right)^{5}\left[2{\bar R}(k_1+k_2)+m(k_1-k_2)\right]^2
 \right]},\nonu
  p(R) \A =\A \displaystyle{64 k_1m \over
 \kappa}\displaystyle{\left[1 \over
 {\bar R}^2\left(1+ \displaystyle{m \over 2{\bar
R}}\right)^{5}\left[2{\bar R}(k_1+k_2)+m(k_1-k_2)\right]^2
 \right]}. \ea
As is clear from Eqs. (31) that the energy momentum tensor
$\tau(R)$ and $p(R)$ have a common singularity, i.e.,  ${\bar
R}\approx 0$. This is due to  the advantage of the isotropic
coordinates. When the geometry is such that it can be interpreted
as a Lorentzian wormhole then the isotropic coordinate patch is a
global coordinate patch. Now one has a single coordinate patch for
the traverse wormhole which will be use to discuss the properties
of the geometry.

Let us explain this result by studying the geometry of each
solution.\\ 1) The geometry of the two solutions (21) and (24) is
invariant under simultaneous sign flip
\[ k_2 \rightarrow -k_2, \qquad \qquad k_1\rightarrow
-k_1,\] it is also invariant under simultaneous inversion
\[ {\bar R} \rightarrow \displaystyle{m^2 \over 4{\cal{\bar R}}}
\qquad and \ \ sign \ \ reversal \qquad \qquad k_2 \rightarrow
-k_2,\] keeping $k_1$ unchanged.\vspace{.3cm}\\
2) \[ k_2\neq0, \qquad \qquad k_1=0,\] gives the
Schwarzschild geometry, it is non-traversable.\vspace{.3cm}\\
 3) \[ k_2=0, \qquad \qquad k_1\neq0,\] gives the
spatial-Schwarzschild traversable wormhole.\vspace{.3cm}\\
4) \[ k_2=0, \qquad \qquad k_1=0\] is a singular.\vspace{.3cm}\\
5)  At the throat $g_{tt}(r=2m)=-k_1^2$, so $k_1\neq0$  is
required to ensure traversability. \vspace{.3cm}\\ 6) To see if
there is ever horizon or not  Dadhich et. al. \cite{DKV} done the
following discussion.

 A horizon would seem to form if the component $g_{tt}=0$, i.e.,
if there is a physically valid solution to \be
k_1(1+\displaystyle{m \over 2{\bar R}})+k_2 (1-\displaystyle{m
\over 2{\bar R}})=0,\ee whose solution has the form \be {\bar
R}_h=\displaystyle{m \over 2}\displaystyle{ k_2 -k_1 \over
k_2+k_1},\ee that is a horizon tries to form if \be \displaystyle{
k_2 -k_1 \over k_2+k_1}>0,\ee which is actually a naked
singularity.  This singularity occurs if either \ba \A \A
k_2+k_1>0, \qquad \qquad and \qquad \qquad k_2-k_1>0, \nonu
\A \A or \nonu
\A \A k_2+k_1<0, \qquad \qquad and \qquad \qquad k_2-k_1<0.\ea
Outside of these regions the naked singularity does not form and
one has a traversable wormhole. Dadhich et. al. \cite{DKV} studied
the $k_1-k_2$ plan and have found that if
\[k_2=Z\cos \theta, \qquad \qquad k_1=Z\sin \theta,\]i)
$\theta=0$: Schwarzschild spacetime.\vspace{.2cm}\\ii) $\theta \in
(0,\pi/4)$, naked singularity.\vspace{.2cm}\\iii)
$\theta=\pi/4\Rightarrow k_2=k_1,$ i.e., the ${\bar R}\rightarrow
0$ region is not flat.\vspace{.2cm}\\ iv) $\theta \in
(\pi/4,3\pi/4)$ traversable wormhole.\vspace{.2cm}\\ v)
$\theta=\pi/2$ is the spatial-Schwarzschild
wormhole.\vspace{.2cm}\\ vi) $\theta=3\pi/4\Rightarrow k_2=-k_1,$
the ${\bar R}\rightarrow \infty$ region is not
flat.\vspace{.2cm}\\vii) $\theta \in (3\pi/4,\pi)$
naked singularity.\vspace{.2cm}\\
viii) $\theta=\pi$: Schwarzschild spacetime.\vspace{.2cm}\\
ix) $\theta>\pi$: repeat the previous treatment.

The  ratio value of the two parameters $k_1$ and $k_2$ that
characterize the solutions (21), (24) is obtained $k_1/k_2 \geq
10^8 \gamma$ with $\gamma^2\beta^2<2$ implying $\beta<\sqrt{2/3}$
\cite{DKV} .

Thus we have two exact solutions of the field equations (11), each
of which leads to the same metric as given by Eq. (22). The axial
vector part of the two solutions (21) and (24) vanishing
identically i.e., $a_{\mu}=0$. Therefore, the two tensors $H_{\mu
\nu}$ and $K_{\mu \nu}$ are also vanishing and M\o ller's tetrad
theory coincides in that case with the teleparallel equivalent of
general relativity.
\newsection{Energy content }

 The superpotential is given by \be {{\cal U}_\mu}^{\nu \lambda} ={(-g)^{1/2} \over
2 \kappa} {P_{\chi \rho \sigma}}^{\tau \nu \lambda}
\left[\phi^\rho g^{\sigma \chi} g_{\mu \tau}
 -\lambda g_{\tau \mu} \gamma^{\chi \rho \sigma}
-(1-2 \lambda) g_{\tau \mu} \gamma^{\sigma \rho \chi}\right], \ee
where ${P_{\chi \rho \sigma}}^{\tau \nu \lambda}$ is \be {P_{\chi
\rho \sigma}}^{\tau \nu \lambda} \stackrel{\rm def.}{=}
{{\delta}_\chi}^\tau {g_{\rho \sigma}}^{\nu \lambda}+
{{\delta}_\rho}^\tau {g_{\sigma \chi}}^{\nu \lambda}-
{{\delta}_\sigma}^\tau {g_{\chi \rho}}^{\nu \lambda} \ee with
${g_{\rho \sigma}}^{\nu \lambda}$ being a tensor defined by \be
{g_{\rho \sigma}}^{\nu \lambda} \stackrel{\rm def.}{=}
{\delta_\rho}^\nu {\delta_\sigma}^\lambda- {\delta_\sigma}^\nu
{\delta_\rho}^\lambda. \ee The energy-momentum density
${\tau_\mu}^\nu$ is defined  by
 \[
 {\tau_\mu}^\nu\stackrel{\rm def.}{=} {{{\cal U}_\mu}^{\nu \lambda}}_{, \
 \lambda},\]
and automatically satisfies the conservation law,
${{\tau_{\mu}}^\nu}_{,\ \nu} =0$. The energy is expressed by the
surface integral \cite{Mo2,MWHL,SNH} \be E=\lim_{r \rightarrow
\infty}\int_{r=constant} {{\cal U}_0}^{0 \alpha} n_\alpha dS, \ee
where $n_\alpha$ is the unit 3-vector normal to the surface
element ${\it dS}$.

Now we are in a position to calculate the energy associated with
solution (21) using the superpotential (36). As is
 clear from (39), the only components which contributes to the energy is ${{\cal U}_0}^{0
 \alpha}$. Thus substituting from solution (21) into
 (36) we obtain the following non-vanishing value
 \be
{{\cal U}_0}^{0 \alpha} ={2 m n^\alpha \over \kappa R^2}.
 \ee
 Substituting from (40) into
(39) we get \be E=m. \ee Repeat the same calculations for solution
(24) we obtain  the necessary components of the superpotential \be
{{\cal U}_0}^{0 \alpha} ={4 m n^\alpha \over \kappa R^2},
 \ee and the energy will have the form
 \be E=2m.\ee
\newsection{Discussion and conculusion}

In this paper we have applied the tetrad having spherical symmetry
with three unknown functions of  radial coordinate \cite{Ro} to
the field equations of M\o ller's tetrad theory of gravitation
\cite{Mo8}. From the resulting partial differential equations we
have  obtained two exact non vacuum solutions.  The solutions in
general are characterize by three parameters $m$, $k_1$ and $k_2$.
If the two parameters $k_1=0$ and $k_2=1$ then one can obtains the
previous solutions when the exponential term equal zero, i.e.,
$e^{-R^3/r1^3}=0$ \cite{Ngr}. The energy-momentum tensor has the
property that $\rho(R)=0$ for the two solutions. This leads to the
violation of the  weak energy and the null energy conditions
defined by Eqs. (26) and (27)  due to the fact that the radial
pressure is negative, i.e., $\tau(R)$ as is clear from Eq. (25).
The line element associated with these solutions has the same form
(22).

To make the picture more clear we discuss the geometry of each
solution. The line element of these solutions in the isotropic
form are given by (29). If $g_{tt}=0$ one obtains a real naked
singularity region. Outside these regions naked singularity does
not form and one obtains a traverse wormhole. The throat of this
wormhole $g_{tt}(R=2m)$ gives the conditions that
$g_{tt}=-{k_1}^2\Rightarrow (k_1\neq0$ is required to ensure the
traversability).  The properties of this wormhole are discussed by
Dadhich et. al. \cite{DKV}.

 It was shown by M\o ller \cite{M2,Mo8} that the tetrad
description of the gravitational field allows a more satisfactory
treatment of the energy-momentum complex than does general
relativity.  We have then applied the superpotential method
\cite{M2,Mo8} to calculate the energy of the gravitating system of
the  two solutions (21) and (24).  As for the first solution we
obtain $E=m$ and there is  no effect of the two parameters $k_1$
 and   $k_2$ characterize the wormhole. As for the second
solution the energy is $E=2m$ and this is due to the fact that the
time-space components of the tetrad fields ${e_0}^\alpha, \
{e_\alpha}^0$ go to zero as $ \displaystyle{1 \over \sqrt{r}}$ at
infinity \cite{SNH,Ngr}. The disappearance of the two parameters
$k_1$ and $k_2$ from the two Eqs. (41) and (43) may give an
impression that these two parameters are not a physical quantities
like the gravitational mass $m$.

\end{document}